\documentclass{PoS}

\title{Using low energy atmospheric neutrinos for precision measurement of the mixing parameters}

\ShortTitle{Low energy atmospheric neutrinos}

\author{\speaker{Hisakazu Minakata}\\ 
        Center for Neutrino Physics, Department of Physics, Virginia Tech, Blacksburg, Virginia 24061, USA \\
        E-mail: \email{minakata71@vt.edu}}

\author{Ivan Martinez-Soler\\
  Theoretical Physics Department, Fermi National Accelerator Laboratory, P.O. Box 500, Batavia IL 60510, USA \&
  Department of Physics and Astronomy, Northwestern University , Evanston, IL 60208, USA \&
  Colegio de F\'{\i}sica Fundamental e Interdisciplinaria de las Am\'{e}ricas (COFI), 254 Norzagaray street, San Juan, Puerto Rico 00901\\
        E-mail: \email{ivan.martinezsoler@northwestern.edu}}

\author{Kimihiro Okumura\\ 
Research Center for Cosmic Neutrinos, Institute for Cosmic Ray Research, University of Tokyo, Kashiwa, Chiba 277-8582, Japan \\ 
       E-mail: \email{okumura@icrr.u-tokyo.ac.jp}}

\abstract{Use of low energy atmospheric neutrinos is considered for precision measurement of neutrino mixing parameters. At around energy $E \simeq$ a few $\times$100 MeV and baseline $L$ of a few $\times$1000 km, CP phase effect is $\sim$10 times larger than that of the conventional LBL accelerator neutrino experiments. We report here a few progresses: (1) To analyze physics in the region, a new perturbative framework at around the solar-scale enhancement is developed. (2) To know the characteristic features of CP $\delta$ dependence of the atmospheric neutrinos at low energies, we plot the ratio of the $\nu_{e}$ and $\bar{\nu}_{e}$ fluxes $F(\delta)/F(\delta=0)$ for $\delta=\pm \frac{\pi}{2}$ and $\pi$ as a function of $E$ using the Honda {\it et al.} flux. Interestingly, it shows $\delta$ dependence of $\simeq$5-10\% level, with positive (negative) sign for $\nu_{e}$ ($\bar{\nu}_{e}$). (3) To reduce the flux systematic errors, measurement of muon energy distribution around $\sim$1 GeV at high altitude may be useful. Water Cherenkov or muon range detectors may be the options with advantage in the latter if it is magnetized in view of the $\delta$ dependence of the flux.
 
 \ }

\FullConference{The 21st international workshop on neutrinos from accelerators (NuFact2019)\\
		August 26 - August 31, 2019\\
		Daegu, Korea}

\begin{document}

\section{Why low energy atmospheric neutrinos?} 

\begin{figure}     
     \includegraphics[width=0.48\textwidth]{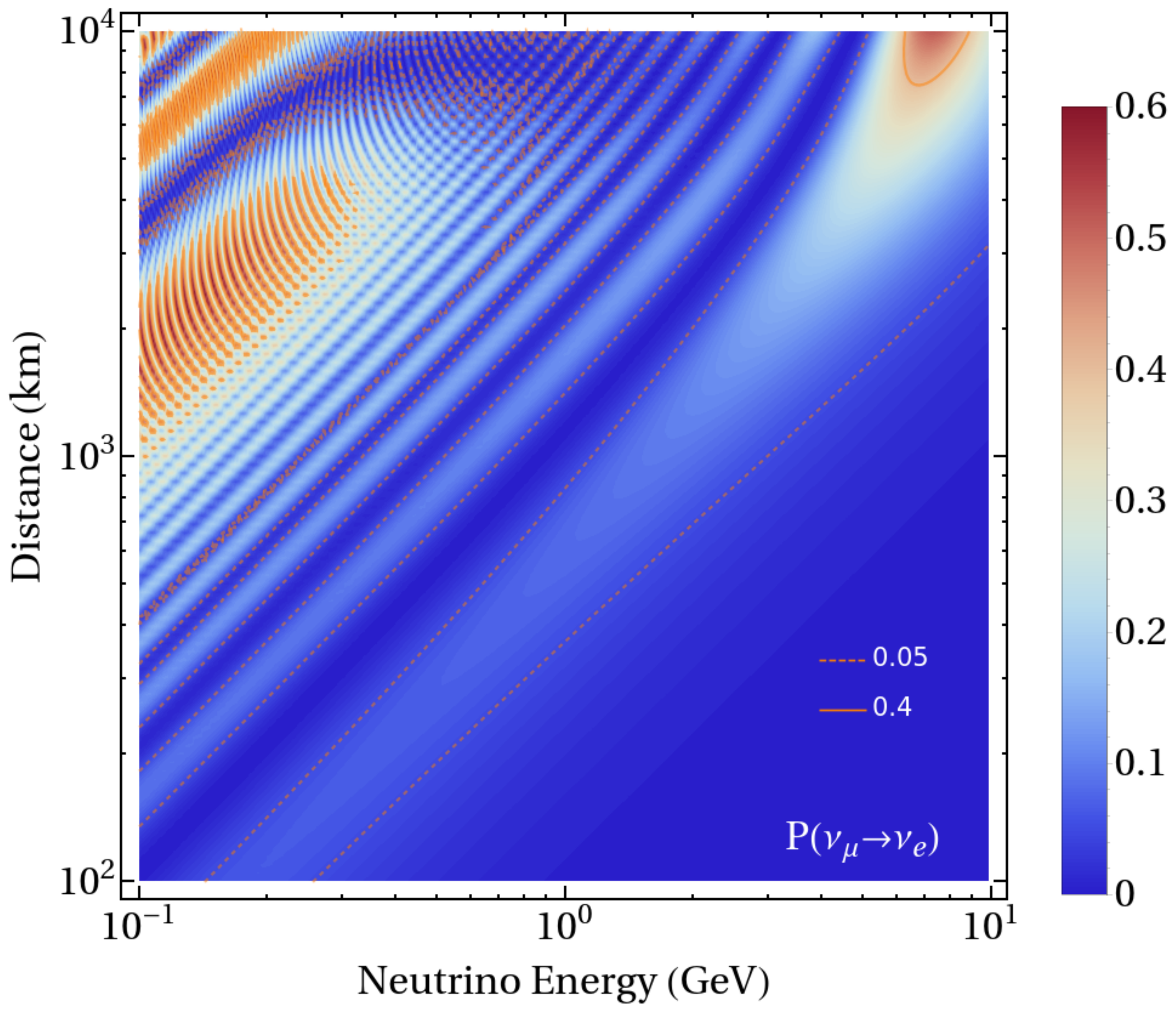}
     \includegraphics[width=0.41\textwidth]{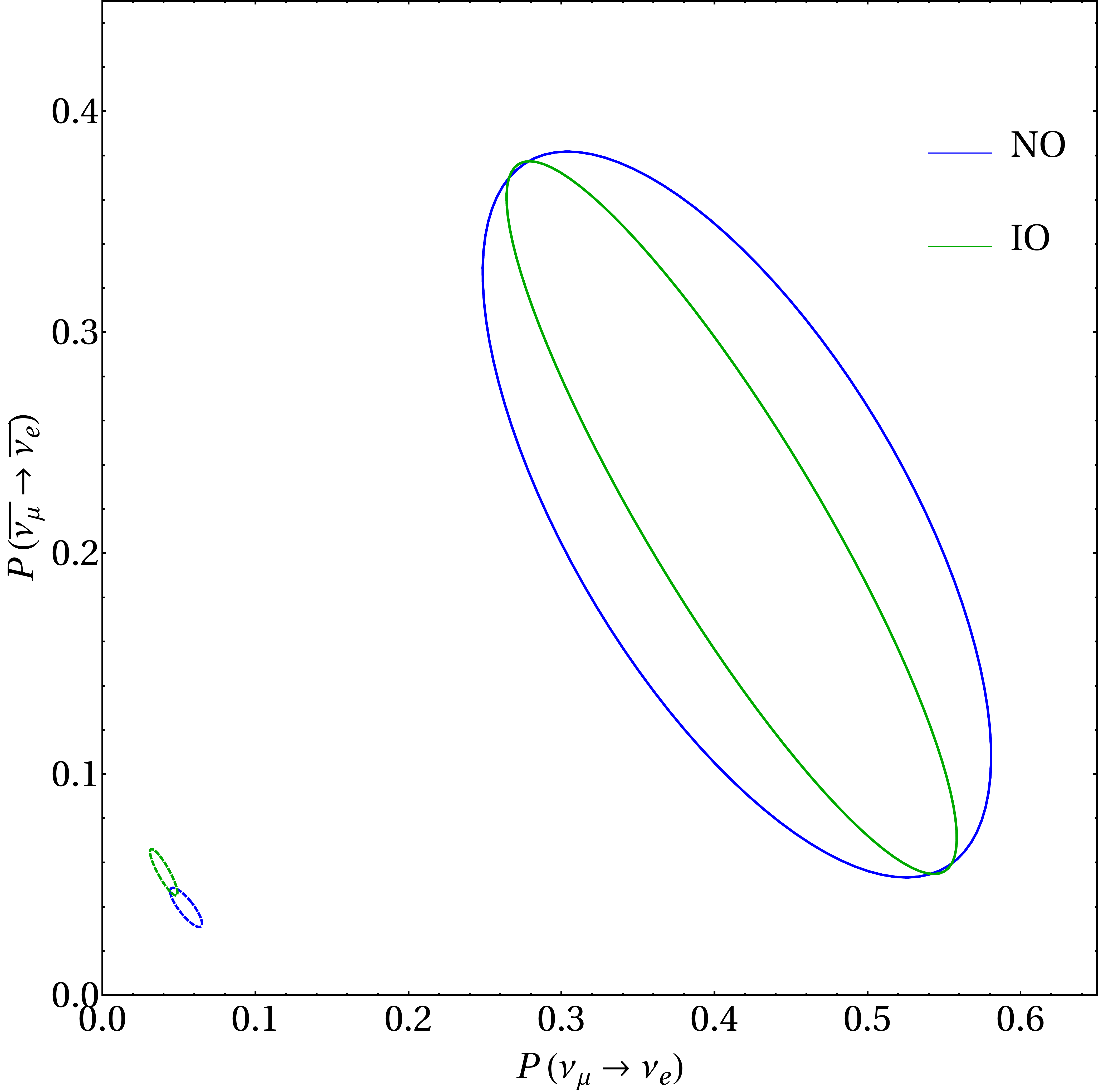}
     \caption{Left panel: The equi-probability contour of $P(\nu_{\mu} \rightarrow \nu_{e})$. Right panel: The bi-probability plot \cite{Minakata:2001qm}, comparing the NO$\nu$A setting (small ellipse) with the region of solar-scale enhancement (large ellipse) \cite{Martinez-Soler:2019nhb}. }
     \label{why-low-E-atm}
     \end{figure}

If we limit ourselves to the terrestrial LBL experiments or atmospheric neutrino observation, there exist only the two regions in the relevant kinematical $E-L$ plane where the appearance probability $P(\nu_{\mu} \rightarrow \nu_{e})$ is large. See the left panel of Fig.~\ref{why-low-E-atm}. Since practically all the accelerator LBL experiments utilize the right-most atmospheric-scale enhanced oscillation region or a constant $L/E$ ``mountain range'' down from it with a few GeV in energy, it is natural to ask why we do not utilize the other one with the solar-scale enhanced region of $P(\nu_{\mu} \rightarrow \nu_{e})$, $E \simeq$ a few $\times 100$ MeV and $L\simeq$ a few $\times 10^{3}$ km \cite{Martinez-Soler:2019nhb}. It is perfectly feasible in atmospheric neutrino experiments to observe the events in such energies and baselines. For the earlier discussions of this subject, see the references in ref.~\cite{Martinez-Soler:2019nhb}.

Then, one may ask: What is the real merit of using data in such low energy region? The simplest answer to this question is: CP phase effect is large. This point can be best represented by using the $P - \bar{P}$ bi-probability plot introduced in \cite{Minakata:2001qm}, in which varying CP phase $\delta$ draws ellipse in $P - \bar{P}$ plane. The size of CP ellipse is $\sim$10 times larger than the NO$\nu$A (as a representative of the LBL expts.) ellipse as one can see in the right panel of Fig.~\ref{why-low-E-atm} \cite{Martinez-Soler:2019nhb}. Physics of much larger effect of $\delta$ is very simple: At around the solar-scale enhanced oscillation one of the suppression factors of CP phase effect, essentially due to the small ratio $\Delta m^2_{21} / \Delta m^2_{31}$, is dynamically lifted \cite{Martinez-Soler:2019nhb}.\footnote{
It may be worth to mention that $\nu_{\mu}$ beam from JPARC extends to lower energy below the peak value of $\simeq600$ MeV. Hence, the detector placed in Korea with the baseline of $\simeq1000$ km is nearby to the right region of the solar-scale enhancement. Therefore, the good CP sensitivity reported in ref.~\cite{Abe:2016ero} may reflect the large CP effect in that region.}

\section{Solar resonance perturbation theory}

To understand the physics in region with the solar-scale enhancement, we have formulated a perturbation theory by which the atmospheric-scale effect can be treated ``perturbatively'' around the solar-scale enhanced oscillations \cite{Martinez-Soler:2019nhb}. But, it must be a peculiar theory as a perturbation theory because the ``perturbed Hamiltonian'' which is of order $\Delta m^2_{31} / 2E$ is $30$ times larger than the solar $2 \times 2$ part of the zeroth order Hamiltonian of order $\Delta m^2_{21} / 2E$. Yet, we have carefully designed the structure of perturbed Hamiltonian in such a way that the $\Delta m^2_{31}$ effect is either decoupled or shows up only in the denominators in the first order (as well as higher order) correction terms. As a consequence, we have a very small effective expansion parameter 
\begin{eqnarray}
A_{ \mbox{exp} } 
&\equiv& 
c_{13} s_{13} 
\biggl | \frac{ a }{ \Delta m^2_{31} } \biggr | 
= 2.78 \times 10^{-3} 
\left(\frac{ \Delta m^2_{31} }{ 2.4 \times 10^{-3}~\mbox{eV}^2}\right)^{-1}
\left(\frac{\rho}{3.0 \,\mbox{g/cm}^3}\right) \left(\frac{E}{200~\mbox{MeV}}\right), 
\label{expansion-parameter}
\end{eqnarray}
where $\rho$ denotes the matter density.

\begin{figure}
\hspace{4mm}
     \includegraphics[width=0.9\textwidth]{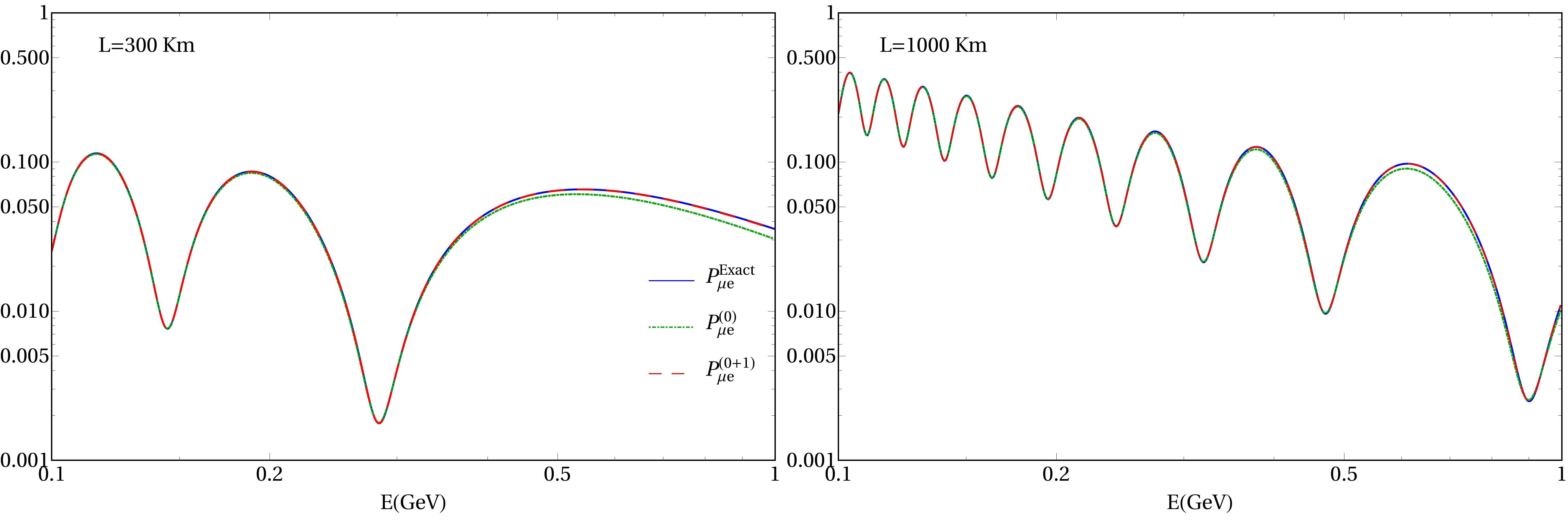} 
     \caption{A comparison between the exact (blue solid), zeroth-order (green dotted), and first-order (red dashed) formulas of $P(\nu_{\mu} \rightarrow \nu_{e})$.} 
 \vspace{-6mm}
     \label{accuracy}
     \end{figure}

Figure \ref{accuracy} demonstrates how accurate is our formula in the $\nu_{\mu} \rightarrow \nu_{e}$ channel. For the $\nu_{\mu} \rightarrow \nu_{\mu}$ channel and the other baselines, see Fig.~3 in ref.~\cite{Martinez-Soler:2019nhb}. It was our surprise to see that the first order formula is accurate even at $L=300$ km, the T2HK baseline, which is much shorter than the one originally designed for the solar resonance perturbation theory. 

\section{Toward precision measurement of atmospheric neutrinos at low energies} 

We propose to utilize the future precision measurement of atmospheric neutrinos at low energies to provide an alternative way to determine accurately the lepton CP phase $\delta$ of the KM type. See ref.~\cite{Martinez-Soler:2019nhb} for the earlier proposals. 

\begin{figure}
\vspace{-1mm}
\hspace{-6mm}
     \includegraphics[width=0.52\textwidth]{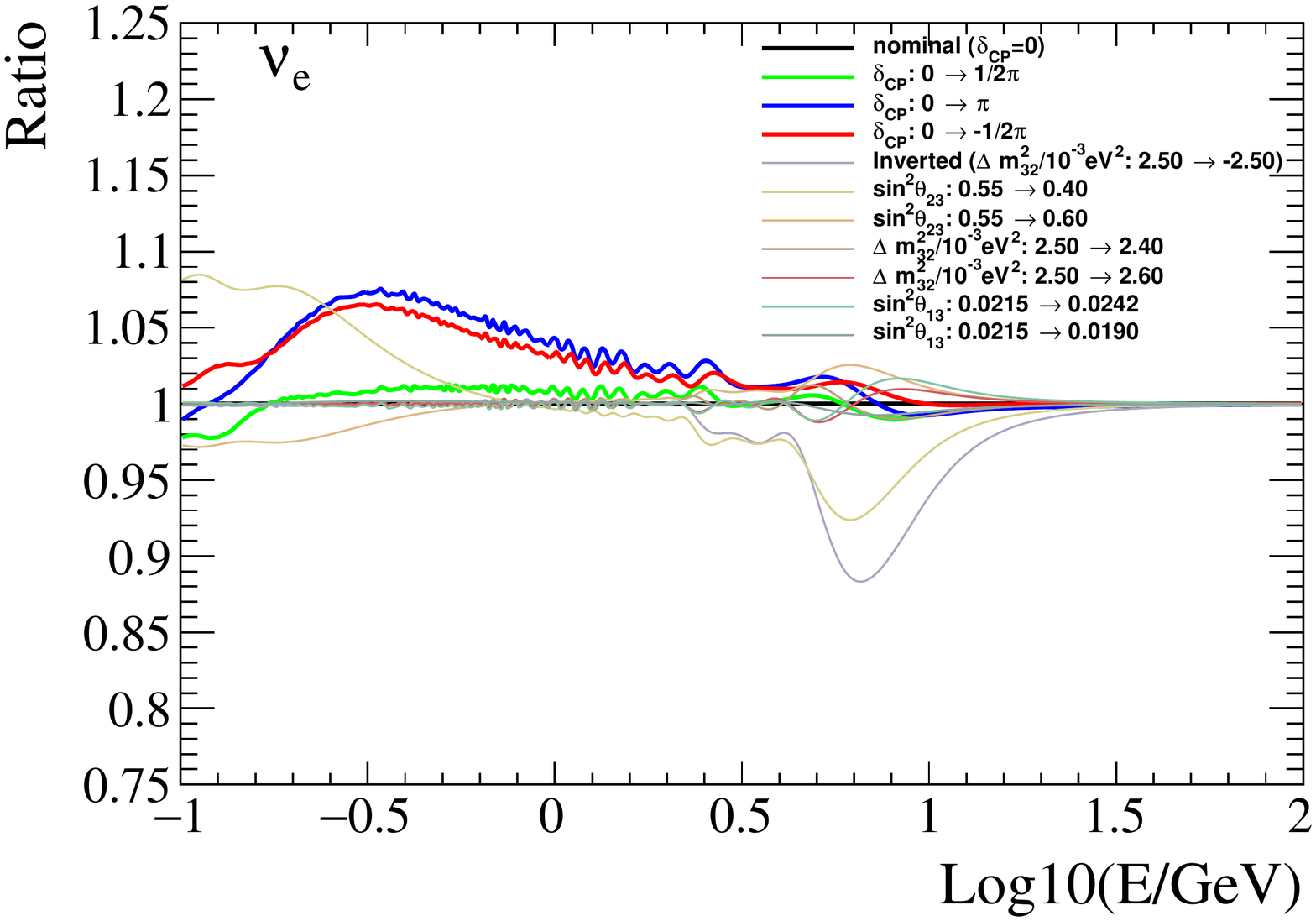}
     \includegraphics[width=0.52\textwidth]{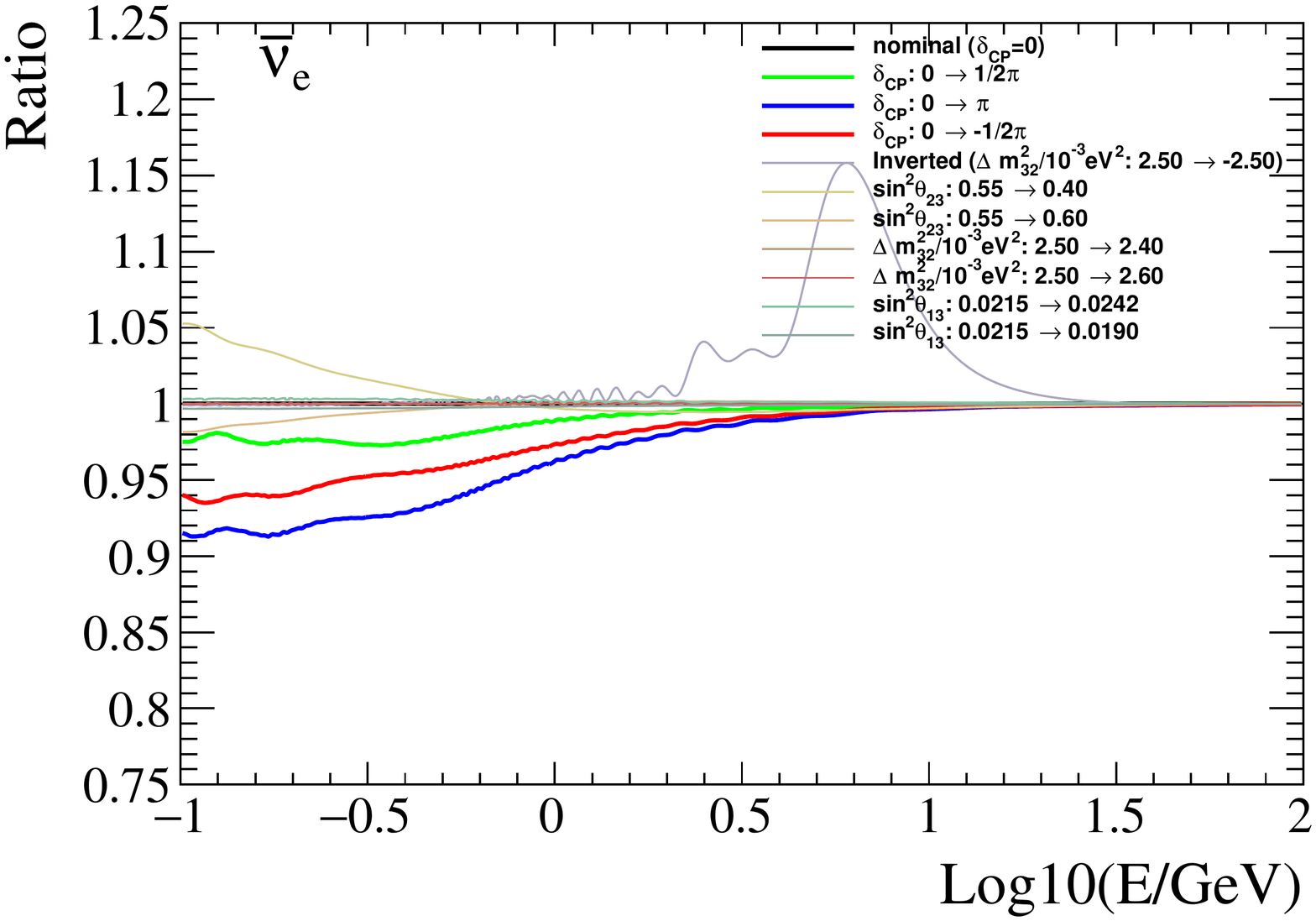}
     \caption{The atmospheric $\nu_{e}$ (left panel) and $\bar{\nu}_{e}$ (right panel)  flux ratios  $F(\delta) / F(\delta=0)$ at Kamioka for the three values of $\delta$, $\pi/2$ (cyan), $\pi$ (blue) and $3\pi/2$ (red) calculated by using eq.~(3.8) in \cite{Richard:2015aua}. }
     \label{okumura-fig}
     \end{figure}

To know how large is the effect of CP phase $\delta$ we plot in Fig.~\ref{okumura-fig} the atmospheric $\nu_{e}$ flux ratio $F(\delta) / F(\delta=0)$ at Kamioka for the three values of $\delta$, $\pi/2$ (cyan), $\pi$ (blue) and $3\pi/2$ (red). The calculation is done by using eq.~(3.8) in \cite{Richard:2015aua} using the HKKM flux \cite{Honda:2011nf}, and by fully implementing the neutrino oscillation. The left panel is for $\nu_{e}$ and the right for $\bar{\nu}_{e}$ fluxes, respectively. For both $\nu_{e}$ and $\bar{\nu}_{e}$ fluxes the effect of CP phase of order $\sim10$\% is expected. A large $\nu-\bar{\nu}$ asymmetry in the ratio $F(\delta) / F(\delta=0)$ suggests that the charge ID of the produced electrons is of key importance. 

\subsection{Which detector do we have and how to improve the systematic errors?}

This is an era of particularly good timing to start thinking about precision measurement of atmospheric neutrinos at low energies. SuperK-Gd will turn on very soon, JUNO will come into operation in less than 2 years, DUNE and Hyper-K will join the allies in 2026-2027. Of particular interest is how good are the performances of the liquid scintillator detector (JUNO) and the liquid Ar TPC (DUNE, see \cite{Kelly:2019itm}) in the context of detection of atmospheric neutrinos at low energies. 

It appears that one of the key issues to make progress is to improve the systematic errors, atmospheric neutrino flux uncertainties and the cross sections. It is discussed \cite{Honda:2019ymh} that measurement of muon energy distribution around $\sim$1 GeV at high altitude, where muon is just born from pion decay, should improve the flux prediction of $\nu_{\mu}$ and $\bar{\nu}_{\mu}$ spectrum. 
Possible ideas for such a detector include (i) water Cherenkov detector, a mini-SK, and (ii) muon range detector used e.g., in the SciBooNE experiment, both at high altitudes. 
The latter, if magnetized, may be advantageous for capability of muon charge ID, but the former could be important to cross check the absolute flux of $\nu_{\mu} + \bar{\nu}_{\mu}$ by a different detector technology. They both may seek a few \% accuracy for the atmospheric muon flux at low energies. 

Finally, we would like to emphasize that to make a real progress the active participation of experimental neutrino community is necessary and is crucial.

\end{document}